\documentclass[preprint,12pt]{elsarticle}

\usepackage{amsmath,amssymb,graphicx,multirow,hyperref,booktabs}
\usepackage[ignoreunlbld,norefs,nocites]{refcheck}
\usepackage{subcaption}
\usepackage{enumitem}
\usepackage[textsize=tiny]{todonotes}
\usepackage[section]{placeins}

\DeclareMathOperator\erf{erf}
\DeclareMathOperator\erfc{erfc}

\DeclareMathOperator\re{Re}

\journal{Computer Physics Communications}

\begin{document}

\begin{frontmatter}

\title{Acceleration of the CASINO quantum Monte Carlo software using
  graphics processing units and OpenACC}

\author[york]{B.\ Thorpe}

\ead{ben.thorpe@york.ac.uk}

\author[york]{M.\ J.\ Smith}

\ead{matthew.j.smith@york.ac.uk}

\author[york]{P.\ J.\ Hasnip}

\ead{phil.hasnip@york.ac.uk}

\author[lancaster]{N.\ D.\ Drummond}

\ead{n.drummond@lancaster.ac.uk}

\address[york]{School of Physics, Engineering \& Technology, University of
    York, York, YO10 5DD, United Kingdom}

\address[lancaster]{Department of Physics, Lancaster
    University, Lancaster, LA1 4YB, United
    Kingdom}

\begin{abstract}
We describe how quantum Monte Carlo calculations using the CASINO
software can be accelerated using graphics processing units (GPUs) and
OpenACC\@.
In particular we consider offloading Ewald summation, the evaluation
of long-range two-body terms in the Jastrow correlation factor, and
the evaluation of orbitals in a blip basis set.
We present results for three- and two-dimensional homogeneous electron
gases and \textit{ab initio} simulations of bulk materials, showing
that significant speedups of up to a factor of 2.5 can be achieved by
the use of GPUs when several hundred particles are included in the
simulations.
The use of single-precision arithmetic can improve the speedup further
without significant detriment to the accuracy of the calculations.

\end{abstract}


\end{frontmatter}


\tableofcontents

\section{Introduction}

Continuum quantum Monte Carlo (QMC) methods are powerful techniques
for finding approximate solutions to the many-body Schr\"{o}dinger
equation.
In these methods the quantum many-body problem is tackled directly in
the continuous, $3N$-dimensional position basis, with Monte Carlo
methods being used to explore and integrate over the high-dimensional
configuration space.
In this article we focus on the variational and diffusion quantum
Monte Carlo (VMC and DMC) methods, which allow one to find the ground
state and selected excited states of fermionic and bosonic systems, as
implemented in the CASINO software
\textcolor{purple}{\cite{Needs_2020,CASINO}}.
A major challenge for these methods is the computational expense,
which not only constrains the range of systems that can be studied,
but also limits the accuracy that can be achieved, as users are forced
to use smaller simulation supercells, larger time steps, etc.
The high cost of QMC methods is partially offset by the fact that QMC
algorithms are inherently parallelizable: one can reduce error bars on
estimated quantities by averaging over independent random walks.
Hence, provided equilibration time is a negligible fraction of total
run time, QMC methods are almost perfect for distributed memory
message-passing interface (MPI) parallelization, and CASINO has been
shown to scale to tens of thousands of MPI processes
\cite{Gillan_2011}.

In addition to distributed memory parallelism, CASINO allows OpenMP
shared memory parallelism in loops over particles in calculations of
the potential energy and calculations of various parts of the
many-body wave function.
This allows one to reduce the walltime for the equilibration stage of
a QMC calculation, and is generally highly efficient when there are
hundreds of particles.
However, efficient parallelism on CPUs is not the only consideration
when examining whether software is suitable for use on modern
high-performance computers; it is also important to assess whether
software can use accelerators such as graphics processing units
(GPUs).
This is the issue that we address in this article.

Since around 2010 the Top 500 list \cite{top500} has been increasingly
dominated by supercomputers featuring GPU accelerators.
At the time of writing the most powerful computer in the world is
\textit{Frontier} at Oak Ridge National Laboratory, which was the
first exascale computer.
Each compute node of Frontier has one 64-core AMD Epyc 7713 ``Trento''
CPU and four 220-core AMD Instinct MI250X GPUs; hence Frontier is
heavily reliant on GPUs to achieve its peak performance.
The growth of machine learning has both driven and been enabled by the
availability of GPU accelerators for training neural networks.
The need to reduce the energy consumption per floating-point operation
has provided a further stimulus to the rise of GPUs in
high-performance computing.

One of the major players in the supercomputing GPU market, the NVIDIA
Corporation, has developed the proprietary compute unified device
architecture (CUDA) platform and application programming interface
(API) \cite{CUDA}.
CUDA allows programmers to have full control over the offloading of
work to NVIDIA GPUs, but is not compatible with the hardware or
compilers of other vendors.
In practice, were a Fortran program such as CASINO to adopt CUDA, it
would be necessary to fork the software into CUDA and non-CUDA
versions, which would pose problems of maintainability and
portability.
Despite the fact that our work focuses on NVIDIA hardware, we are keen
to avoid these challenges associated with the use of CUDA, and we have
adopted solutions that allow us to retain a single code base as well
as the support of a plurality of compiler and hardware vendors.

A directives-based approach provides a viable technique for allowing
control of offloading from within a single Fortran code base.
The OpenACC programming standard and API \cite{OpenACC} is supported
by Fortran compilers including NVIDIA's nvfortran, the Cray/HPE
Fortran compiler, and GNU's gfortran.
OpenACC is developed and overseen by the OpenACC Organization, a
non-profit organization whose members include the NVIDIA Corporation,
AMD, HPE, ARM, and Oak Ridge National Laboratory (home to the Frontier
supercomputer), as well as many \textcolor{purple}{high-performance
  computing} centers and academic institutions.
\textcolor{purple}{OpenACC fully supports Fortran, C, and C++\@.}
OpenACC compiler directives are used to identify parts of a program
that should be compiled for execution on an accelerator device;
furthermore these directives are also used to provide much of the
functionality specific to programming for GPUs.

These directives include: 
\begin{itemize}
  \item DATA directives, to control the transfer of data between the
    main memory and the device memory.
  \item ASYNC and WAIT directives, to allow asynchronous kernel
    operations for overlapping computation with data movement.
  \item GANG, WORKER, and VECTOR directives, to enable developers to
    map multidimensional arrays and nested loops more directly to the
    available hardware without relying on the compiler.
\end{itemize}

A second directives-based approach is the extension of the OpenMP API
\cite{OpenMP} to support offloading to generic accelerator devices,
including GPUs.
OpenMP is developed by the OpenMP Architecture Review Board, which is
supported by a broad range of vendors in the high-performance
computing sector.
OpenMP provides a well-established and widely used standard for
handling multiple CPU threads in numerical software.
As with OpenACC, offloading to accelerators is controlled by means of
compiler directives, to compile specific sections of the program for
execution on an accelerator.
These also allow the developer to control data movement between the
device and the host CPU, as well as performing asynchronous operations
and a limited amount of nested/multilevel parallelism through the use
of single-instruction, multiple-data (SIMD) teams.

From a developer's point of view, the existence of similar, but
incompatible, competing standards, each with varying levels of support
from the different vendors, is not ideal.
In this work we have chosen to use OpenACC, primarily because the
NVIDIA developer ecosystem is the most mature, and currently has
better support for the OpenACC API than OpenMP\@.
Furthermore, OpenACC is specifically designed with accelerators in
mind and allows the programmer and the compiler more scope to optimize
the directives to make best use of the underlying hardware.
If in the future OpenMP should emerge as the dominant standard, it
will be relatively easy for us to adapt our OpenACC implementation to
OpenMP, as OpenMP is already in use within CASINO for CPU
parallelism\@.

Other QMC codes have each taken a different approach for supporting
GPU computing.
In the early 2010s, the QMCPACK code \cite{Kent_2020} forked into a
limited-feature CUDA version and a full-feature non-CUDA version, but
QMCPACK has more recently adopted OpenMP for offloading from a single
code base.
The PyQMC module \cite{Wheeler_2023} in PySCF \cite{Sun_2024} makes
use of the CuPy library \cite{CuPy}, which in turn uses CUDA and AMD's
ROCm software stack for GPU computing \cite{ROCm}, to allow the use of
GPUs to complete various tasks.
The European version of the CHAMP code \cite{Champ-EU} can make use of
NVIDIA's cuBLAS \cite{cuBLAS} library for linear algebra, while the
North American version of CHAMP \cite{CHAMP-US} supports GPUs via the
einspline library for spline representation of parts of the wave
function \cite{einspline}.
The TurboRVB QMC code uses OpenMP for offloading \cite{Nakano_2020}.
To our knowledge, apart from an early experiment \cite{Anderson_2007}
with the QMcBeaver code \cite{Kent_2007}, OpenACC has not been used in
a major QMC code.

The rest of this paper is structured as follows.
In Section \ref{sec:targets} we describe the aspects of a CASINO
calculation that are particularly promising for offloading to GPUs.
The tools used for profiling CASINO on CPUs and GPUs are discussed in
Section \ref{sec:profiling}.
In Section \ref{sec:quantifying_GPU_perf} we take a small but useful
diversion into how exactly one can quantify GPU performance.
In Section \ref{sec:openacc_casino} we describe our modifications to
CASINO to allow the use of GPUs.
The use of single precision (SP) arithmetic rather than double
precision (DP) arithmetic is discussed in Section \ref{sec:sp}.
We present performance results in Section \ref{sec:results}.
Finally, we draw our conclusions in Section \ref{sec:conclusions}.
Throughout we use Hartree atomic units ($\hbar=m_{\rm
  e}=|e|=4\pi\epsilon_0=1$ a.u.), except where otherwise stated.

\section{Targets in CASINO for offloading to GPUs \label{sec:targets}}

\subsection{VMC and DMC methods}

In the VMC method we use the Metropolis algorithm to sample particle
configurations distributed as the square modulus of a trial spatial
wave function $\Psi$, and we average $(\hat{A}\Psi)/\Psi$ over these
configurations to obtain an unbiased estimate of $\langle \Psi |
\hat{A} | \Psi \rangle/\langle \Psi|\Psi \rangle$, where $\hat{A}$ is
a spin-independent quantum operator \cite{Foulkes_2001}.
Free parameters in the trial wave function are optimized by minimizing
either the energy expectation value or the variance of the energy.

In the DMC method \cite{Foulkes_2001,Ceperley_1980} the ground-state
component of the trial wave function is projected out by simulating
drift, diffusion, and branching/dying processes governed by the
time-dependent Schr\"{o}dinger equation in imaginary time.
For the ground state of a bosonic system there are no uncontrolled
approximations; for fermionic systems or for excited states we solve
the imaginary-time-dependent Schr\"{o}dinger subject to the boundary
condition that the wave function remains fixed at zero at the nodes of
the trial wave function $\Psi$.
Expectation values are computed in a similar manner to VMC\@.

In practice the VMC and DMC methods both involve repeatedly proposing
single-particle moves, calculating the value and first and second
derivatives of the trial wave function $\Psi$, and calculating
(changes in) the potential energy.

The trial wave functions used in QMC calculations are typically of
Slater-Jastrow-backflow form
\begin{equation} \Psi({\bf R})=e^{J({\bf R})} S({\bf
  X}({\bf R})), \end{equation}
where $S({\bf R})$ is an expansion of one or more Slater determinants
of single-particle orbitals, $e^{J({\bf R})}$ is a Jastrow correlation
factor and ${\bf X}({\bf R})$ is a backflow transformation of the
$3N$-dimensional vector of particle coordinates ${\bf R}$.
The orbitals in the Slater determinant are usually taken from either
density functional theory (DFT) calculations or wave-function-based
quantum chemistry calculations, e.g., Hartree-Fock theory.
In this work we will focus on single-determinant Slater-Jastrow wave
functions without a backflow transformation [i.e., we will assume that
  ${\bf X}({\bf R})={\bf R}$ and that $S$ is the product of a Slater
  determinant of occupied orbitals for spin-up electrons and a Slater
  determinant of occupied orbitals for spin-down electrons].
CASINO's standard Jastrow factor \cite{Drummond_2004} is described in
Section \ref{sec:lrjas}.

Each time step of a Slater-Jastrow QMC simulation of $N$ particles
requires the evaluation of $O(N^2)$ long-range pairwise Jastrow terms,
each of $O(N)$ orbitals expanded in a localized basis set at each of
the $N$ particle positions, and the sum of the $O(N^2)$ pairwise
Coulomb interactions; the cost hence scales as $O(N^2)$.
The cost of updating the Slater determinants is $O(N^3)$ but with a
sufficiently small prefactor that this is only relevant at the very
largest system sizes that have been studied using QMC (1000--2000
particles).
As an aside, we note that the number of time steps required to achieve
a given statistical error bar must grow as the system size to
counteract the growth in the variance, and hence the scaling of
standard Slater-Jastrow QMC is $O(N^3+\epsilon N^4)$
\cite{Foulkes_2001}.

\subsection{Ewald summation}

In a molecular system, the electron-electron Coulomb potential energy
is simply a sum of pairwise $1/r$ interactions.
Even with thousands of electrons, this is cheap to evaluate.
On the other hand, to model bulk condensed matter, we study simulation
supercells subject to periodic boundary conditions and hence we need
to use periodic Coulomb potentials such that the electrostatic energy
tends to a well-defined and appropriate value in the limit of infinite
simulation-cell size.

The most well-known and widely used method for evaluating Coulomb
interactions in periodic (crystalline) systems is Ewald summation
\cite{Ewald_1921}.
The Coulomb potential is taken to be the periodic solution to
Poisson's equation, which is evaluated by means of rapidly convergent
series in real space and reciprocal space.
Physically, the use of the periodic solution of Poisson's equation
corresponds to the absence of a macroscopic polarization field
throughout the crystal, as would be the case if the macroscopic
crystal were wrapped in a perfect conductor so that surface
polarization charges were cancelled by image charges in the conductor.
The electrostatic energy of a set of charges $\{q_i\}$ with positions
$\{{\bf r}_i\}$ in a periodic simulation cell is
\begin{equation} \hat{V}_{\rm Ew} = \sum_{i=1}^{N-1} \sum_{j=i+1}^N q_i q_j
v_{\rm E}({\bf r}_i-{\bf r}_j) + \frac{1}{2}\sum_j q_j^2 v_{\rm M},
  \end{equation}
where
\begin{eqnarray} v_{\rm E}({\bf r}) & = & \frac{4 \pi}{\Omega_{\rm s}}
\sum_{{\bf G}_{\rm s} \neq {\bf 0}} \frac{e^{-G_{\rm s}^2/(4\gamma) +
    i{\bf G}_{\rm s} \cdot {\bf r}}}{G_{\rm s}^2} +\sum_{{\bf R}_{\rm
    s}} \frac{ \erfc \left(\sqrt{\gamma}|{\bf r}-{\bf R}_{\rm
    s}|\right)}{|{\bf r}-{\bf R}_{\rm s}|} - \frac{\pi}{\Omega_{\rm s}
  \gamma} \label{eq:Ewald} \\ v_{\rm M} & = & \frac{4 \pi}{\Omega_{\rm
    s}} \sum_{{\bf G}_{\rm s}\neq {\bf 0}} \frac{e^{-G_{\rm
      s}^2/(4\gamma)}}{G_{\rm s}^2}+\sum_{{\bf R}_{\rm s}\neq{\bf 0}}
\frac{\erfc \left(\sqrt{\gamma}R_{\rm s}\right)}{R_{\rm s}} -2
\sqrt{\frac{\gamma}{\pi}}-\frac{\pi}{\Omega_{\rm s} \gamma},
    \end{eqnarray}
and $\Omega_{\rm s}$ is the volume of the simulation cell, $\{{\bf
  R}_{\rm s}\}$ are the simulation-cell lattice points, $\{{\bf
  G}_{\rm s}\}$ are the simulation-cell reciprocal lattice points, and
$\gamma$ is a constant that can be chosen to minimize the cost of
evaluating Eq.\ (\ref{eq:Ewald}) \cite{Ewald_1921}.
\textcolor{blue}{Similar expressions can be derived for Ewald
  interactions in two dimensions
  \cite{Parry_1975,Parry_1976,Wood_2004}:
\begin{eqnarray}
v_{\rm E}({\bf r}) & = & \frac{\pi}{A_{\rm s}} \sum_{{\bf G}_{\rm s}
  \neq {\bf 0}} \frac{e^{zG_{\rm s}} \erfc \left(\frac{G_{\rm
      s}}{2\sqrt{\gamma}}+z\sqrt{\gamma}\right)+e^{-z G_{\rm s}} \erfc
  \left(\frac{G_{\rm s}}{2\sqrt{\gamma}}-z\sqrt{\gamma}\right)}{G_{\rm
    s}} e^{i {\bf G}_{\rm s} \cdot {\bf r}} \nonumber \\ & & {} +
\sum_{{\bf R}_{\rm s}} \frac{\erfc \big(\sqrt{\gamma}|{\bf r} - {\bf
    R}_{\rm s}|\big)}{|{\bf r} - {\bf R}_{\rm s}|} -
\frac{2\pi}{A_{\rm s}} \left[ z \erf (z\sqrt{\gamma}) +\frac{e^{-\gamma
      z^2}}{{(\gamma\pi)}^{\frac{1}{2}}}\right] \label{eq:Ewald2D} \\ v_{\rm M}
& = & \frac{2\pi}{A_{\rm s}}\sum_{{\bf G}_{\rm s} \neq {\bf
    0}}\frac{\erfc \left(\frac{G_{\rm
      s}}{2\sqrt{\gamma}}\right)}{G_{\rm s}} + \sum_{{\bf R}_{\rm s}
  \neq {\bf 0}} \frac{\erfc \left(\sqrt{\gamma} R_{\rm
    s}\right)}{R_{\rm s}} - \frac{2\sqrt{\gamma}}{\pi^{\frac{1}{2}}} -
\frac{2 \pi^{\frac{1}{2}}}{A_{\rm s}\sqrt{\gamma}},
\end{eqnarray}
where $A_{\rm s}$ is the area of the simulation cell.}

It is clear that the cost of evaluating the Ewald interaction in
Eqs.\ (\ref{eq:Ewald}) or (\ref{eq:Ewald2D}) is vastly greater than the
cost of evaluating $1/r$.
\textcolor{blue}{Equations (\ref{eq:Ewald}) and (\ref{eq:Ewald2D}) are
  suitable for offloading to GPUs because the evaluation of the series
  is dominated by floating point operations and both the real- and
  reciprocal-space series are inherently parallelizable; moreover,
  much of the data required can be moved to the GPU once, at the
  outset of the calculation.
Offloading of Ewald interactions is investigated in Section
\ref{sec:OpenACC_Ewald}.}

\subsection{Long-range Jastrow terms \label{sec:lrjas}}

CASINO's standard Jastrow exponent is of the form
\begin{eqnarray} J  & = & \sum_{i=1}^N \left[ \sum_{I=1}^{N_{\rm n}} \chi(r_{iI})
+ q({\bf r}_i) \right] + \sum_{i=1}^{N-1} \sum_{j=i+1}^N \left[
    u(r_{ij}) + p({\bf r}_{ij}) \right] \nonumber \\ & & {} +
  \sum_{i=1}^{N-1} \sum_{j=i+1}^N \sum_{I=1}^{N_{\rm n}}
  f(r_{iI},r_{jI},r_{ij}) + \sum_{i=1}^{N-2} \sum_{j=i+1}^{N-1}
  \sum_{k=j+1}^N H(r_{jk},r_{ki},r_{ij}), \label{eq:J} \end{eqnarray}
where $\chi(r)$, $u(r)$, $f(r,r',r'')$, and $H(r,r',r'')$ are
polynomial functions of interparticle distances, and $p({\bf r})$ and
$q({\bf r})$ are plane-wave expansions in the reciprocal lattice
points of the simulation cell and the primitive cell, respectively
\cite{Drummond_2004,Lopez_2012}.
$N_{\rm n}$ is the number of nuclei, ${\bf r}_i$ is the position of
particle $i$, ${\bf r}_{ij}$ is the nearest-image displacement between
particles $i$ and $j$, and $r_{iI}$ is the nearest-image distance
between particle $i$ and nucleus $I$.
The functions $u$, $\chi$, etc., may depend on the species of particle
involved (e.g., $u$ is nearly always different for parallel- and
antiparallel-spin electrons due to differences in the Kato cusp
conditions) and, where relevant, the functions depend on the ions
involved.

The four terms in Eq.\ (\ref{eq:J}) are, respectively, an
inhomogeneous one-body term, a homogeneous two-body term, an
inhomogeneous two-body term, and a homogeneous three-body term.
The Jastrow factor does not affect the wave-function (anti)symmetry
described by the Slater part of the wave function.
The Jastrow factor is nodeless and hence allows a description of
correlation without altering the excitation level described by the
Slater wave function.
The polynomial two-body term $u(r)$ is used to enforce the
electron-electron Kato cusp conditions \cite{Kato_1957}, and the
polynomial one-body term $\chi(r)$ can be used to enforce the
electron-nucleus cusp conditions.
The other terms are required to be at least twice differentiable
everywhere.

The homogeneous two-body correlations $u$ and $p$ describe both
short-range dynamical correlations (the Kato cusp behavior) and
long-range behavior that is well-described by the random phase
approximation \cite{Foulkes_2001}.
The inhomogeneous one-body terms $\chi$ and $q$ are primarily required
to undo the effects of two-body correlations on the charge density
around atoms described by the Slater wave function; hence they are
mostly short-ranged.
The inhomogeneous two-body correlations $f$ are intended to describe
the density dependence of two-body correlations inside atoms, and
hence are primarily short-ranged.
The homogeneous three-body term $H$ is rarely used except in model
systems such as electron-hole complexes \cite{Mostaani_2017} or cold
atomic gases \cite{Drummond_2011}.
Hence the cost of evaluating the Jastrow factor is usually dominated
by the evaluation of the homogeneous two-body terms.
We therefore focus on offloading the evaluation of the values and
derivatives of the two-body terms $u(r)$ and $p({\bf r})$.

The forms of $u$ and $p$ are as follows \cite{Drummond_2004}:
\begin{eqnarray} u(r) & = & \sum_{l=0}^{N_u} \alpha_l r^l (r-L_u)^C
\Theta(L_u-r) \label{eq:u} \\ p({\bf r}) & = & \sum_{m=1}^{N_p} a_m
\sum_{{\bf G}_{\rm s}\in \bigstar_m^+} \cos \left( {\bf G}_{\rm s}
\cdot {\bf r} \right), \label{eq:p} \end{eqnarray}
where $\{\alpha_i\}$ are optimizable parameters for $i=0$ and
$i=2,\ldots,N_u$, while $\alpha_1$ is determined by the Kato cusp
conditions, $L_u$ is a cutoff length that must be less than half the
distance from the origin to the closest nonzero simulation-cell
lattice point, $\Theta$ is the Heaviside function and $C$ is the
number of derivatives of $u(r)$ that are well-defined at $r=L_u$;
usually we have $C=2$ or $C=3$.
Typically the expansion order $N_u$ is chosen to be in the range
8--12.
The $\{a_m\}$ are optimizable coefficients for each star $\bigstar_m$
of simulation-cell reciprocal-lattice points ${\bf G}_{\rm s}$; the
``+'' superscript indicates that only one ${\bf G}_{\rm s}$ vector out
of each $\pm{\bf G}_{\rm s}$ pair is included in the sum.
Typically $N_p=3$--$15$ stars of ${\bf G}_{\rm s}$ vectors are used.

For ${\bf G}_{\rm s}=n_1{\bf b}_1 + n_2{\bf b}_2 + n_3{\bf b}_3$,
where $\{ {\bf b}_i \}$ are supercell reciprocal lattice vectors, we
may write
\begin{equation} \cos({\bf G}_{\rm s} \cdot
  {\bf r}_{ij}) = \re\left[ {(e^{i{\bf b}_1\cdot{\bf r}_{ij}})}^{n_1}
    \, {(e^{i{\bf b}_2\cdot{\bf r}_{ij}})}^{n_2} \, {(e^{i{\bf
          b}_3\cdot{\bf r}_{ij}})}^{n_3}
    \right], \label{eq:cosp}\end{equation}
so only three complex exponentials are required to compute all the
required cosines for Eq.\ (\ref{eq:p}).
The powers of these exponentials are computed and buffered.
The required number of reciprocal lattice points is typically
$\sim$10--$\sim$100.
We note that for large plane-wave expansions there is scope to reduce
the computational cost of the $p$ term by means of a blip (B-spline)
re-representation.

The modifications to CASINO to enable offloading of the long-range two
body Jastrow terms are described in Section \ref{sec:u_and_p}.

\subsection{Evaluation of orbitals in a blip basis set}

Especially in studies of condensed matter, it is common practice to
use a plane-wave basis set to represent single-particle orbitals in
DFT calculations.
However, QMC simulations are performed in real space, and hence a
plane-wave basis set is a particularly expensive choice: after each
single-particle move, all $O(N)$ plane-wave basis functions must be
evaluated at the particle's new position; evaluation of the $O(N)$
occupied orbitals at the new position requires $O(N^2)$
multiplications of basis functions by orbital coefficients.
To reduce the cost we may re-represent the orbitals in a localized
blip (B-spline) basis set before the start of a QMC calculation
\cite{Williamson_2001,Alfe_2004}.
The cost of evaluating the orbitals after each particle move is then
reduced to the cost of looking up the particle's position in the
spline grid followed by the evaluation of a small number of cubic
polynomials defining the blip basis functions, followed by
multiplication of the blip functions by coefficients for each orbital
[$O(N)$ multiplications].

We investigate offloading the multiplication of the blip basis
functions by the blip coefficients for each orbital in
Section \ref{sec:blips}.
A similar approach could be taken for evaluating orbitals in other
basis sets, e.g., Gaussian basis sets.

\section{Profile-driven GPU porting of CASINO \label{sec:profiling}}

\subsection{Development cycle}

The development cycle for porting CASINO to GPUs using OpenACC is
outlined as follows.
\begin{enumerate}[label=(\arabic*)]
    \item Profile the application to identify performance bottlenecks.
    \item Assess which of the bottlenecks identified in (1) are
      amenable to acceleration by GPUs; refactor CPU code as necessary
      to expose parallelism maximally.
    \item Use OpenACC to parallelize (or to offload to an appropriate
      library) a bottleneck prioritized from those in (2).
    \item Minimize host-device data migrations resulting from (3).
    \item Optimize the directives introduced in (3) and (4) for given
      architectures and/or inputs.
\end{enumerate}

The cycle is executed incrementally to ensure correctness, and
iteratively to achieve the maximum performance improvement.

\subsection{Subroutine timers}

Profiling CASINO is made straightforward by an internal timer system,
which allows the walltime, CPU time, and number of invocations for
different activities to be recorded and reported at the end of a
calculation.
Extensive profiling of CASINO as part of the European Union's
Performance Optimization and Productivity (POP) program \cite{pop}
yielded results in good agreement with the results of CASINO's
internal timers.

CASINO supports a very broad range of use cases, from studies of
molecules with compact Gaussian or Slater-type-orbital basis sets and
multideterminant Slater-Jastrow(-backflow) wave functions, to studies
of nanomaterials and condensed matter using blip basis sets and
Slater-Jastrow(-backflow) wave functions, to studies of plasmas of
particles with a range of different interactions moving in one, two,
or three dimensions with Slater-Jastrow(-backflow) or pairing wave
functions.
As a representative example of the breakdown of timings for a large
\textit{ab initio} condensed matter system, we consider a DMC
calculation in a $3 \times 3 \times 3$ supercell of BaTiO$_3$, with a
Slater-Jastrow wave function and a blip basis set for the orbitals.
The atomic cores are described by Dirac-Fock pseudopotentials
\cite{Trail_2005a,Trail_2005b}, such that the supercell contains 648
electrons.
A target population of 512 walkers is used, together with a time step
of 0.01 Ha$^{-1}$.
In addition to the total energy, the DMC charge density is
accumulated.
For the baseline CPU timings and profile, CASINO was compiled using
the Intel Fortran compiler and run on a 64-core Intel Ice Lake node.
CPU timings broken down by activity are shown in Table
\ref{table:batio3_timings}.
Activities not accounted for in Table \ref{table:batio3_timings}
include: the updating and copying of arrays when particle moves are
proposed and accepted, and after branching of walkers; evaluating
random numbers; checkpointing; etc.
Summing energies over MPI processes is expensive because it is a
blocking operation, and the populations of walkers assigned to each
process fluctuate due to branching.

\begin{table}[!htbp]
\centering
\caption{Percentage of CPU time spent on different activities in a
  648-electron BaTiO$_3$ DMC calculation with a Slater-Jastrow trial
  wave function. \label{table:batio3_timings}}
\begin{tabular}{lc}
\toprule

Activity & Percentage of CPU time \\

\midrule

Evaluating Jastrow $u$ terms & 1.7\% \\

Evaluating Jastrow $\chi$ terms & 0.5\% \\

Evaluating Jastrow $f$ terms & 10.4\% \\

Evaluating Jastrow $p$ terms & 51.5\% \\

Evaluating e-e and e-n distances & 1.1\% \\

Evaluating local e-e and e-n potentials & 1.8\% \\

Evaluating blip orbitals & 7.9\% \\

Updating Slater cofactor matrix & 1.4\% \\

Summing energies over MPI processes & 16.0\% \\

Charge density accumulation & 1.5\% \\

\bottomrule
\end{tabular}
\end{table}

\subsection{Profiling performance on GPUs}

Having identified the CPU performance bottlenecks, different aspects
of the simulation were prioritized in order of their respective
percentages of CPU time as reported in Table
\ref{table:batio3_timings}.
Vital to the success of the development cycle at this stage is
repeated use of GPU profiling tools such as NVIDIA's Nsight Systems
\cite{nvidia_nsight_systems} and Nsight Compute
\cite{nvidia_nsight_compute}, and AMD's ROCProfiler
\cite{amd_rocprofiler} and Omniperf \cite{amd_omniperf}.
Each of these tools is incompatible with hardware other than that of
its particular vendor; NVIDIA's Nsight ecosystem was used extensively
in this work since our primary focus is on using OpenACC to port to
this vendor's GPUs.

Nsight Systems provides system-wide performance traces and metrics; a
Systems profile spans the CPU, GPU, and memory, thereby facilitating
the correlation of performance inhibitors across host and device.
The main interface is a graphical timeline representing the database
of samples collected during a given run time, and derived metrics.
We typically used this timeline in the early stages of the GPU
development cycle to obtain a bird's-eye view of host-device memory
copies and CUDA API calls resulting from the initial decoration of
targeted code with OpenACC directives.
Problematic areas were subsequently analyzed in more detail using
Nsight Compute.

Nsight Compute presents detailed performance metrics obtained via
hardware counters and software instrumentation, exposing the low-level
GPU processes with which CUDA interfaces on a per-kernel level.
It then uses a predefined set of rules to provide a guided analysis of
these metrics to the user.
Many of the metrics are correlated to lines of source code, and
additionally to the assembly generated by the compiler; we could thus
navigate directly to relevant lines in CASINO and gain insight into
the mechanics underlying observed performance.
These insights not only informed optimization of the directly affected
source code, but also improved the efficiency of our development by
preemptively mitigating GPU performance inhibitors when subsequently
coding during the iterative porting process.

One of the main utilities of Nsight Compute in our development cycle
was to examine GPU throughput (i.e., NVIDIA's \textit{Speed Of Light}
metrics), as the achieved percentage of utilization with respect to
the theoretical maximum, for both compute [i.e., the GPU's streaming
multiprocessors (SM); see Section \ref{sec:quantifying_GPU_perf} for
more details] and memory; the limitations to device occupancy due to
respective usage of registers and shared memory emerged as metrics of
particular interest (see Section \ref{sec:OpenACC_Ewald}).
Nsight Compute's \textit{Launch Statistics} provided details of SM and
memory utilization per OpenACC parallel region, reporting the GPU
resources assigned at run time to a given kernel.
We used these statistics to inform experimentation with the OpenACC
\texttt{num\_gangs}, \texttt{num\_workers}, and
\texttt{vector\_length} clauses to optimize parallel regions.

Subsequent to each optimization, a new profile was generated in order
that any impact on performance could be assessed.
Nsight Compute facilitates this iterative process as it can use a
previously generated report as a baseline, automatically calculating
any difference in metrics between this and those in the new profile.

\section{Understanding and quantifying GPU performance
\label{sec:quantifying_GPU_perf}}

Modern GPUs typically comprise thousands of small, highly parallel
cores and are capable of processing thousands or even millions of
instructions simultaneously.
The cores of a GPU are typically low-powered relative to those of a
CPU, and are also designed to perform quite different tasks.
Direct comparison of the performance of the former with the latter is
therefore not always enlightening.
Thus, to quantify better the theoretical maximum performance of the
GPU it may help to take a small diversion into how GPUs work at a
hardware level.

The exact naming conventions and specifics vary across different
hardware vendors and even between devices from the same manufacturer.
However, in general, GPUs consist of a large number of processing
elements (compute clusters), each of which contains a number of GPU
cores and some scheduling hardware to distribute a queue of
instructions over a number of fixed-size pools of threads.
Each GPU core runs the same instruction across a pool of threads 
in lockstep using so-called vector processors.
The GPU cores themselves generally consist of a number of 16-, 32-,
and 64-bit integer and floating-point vector processors and some
arithmetic logic units (ALUs).
Each cluster also contains a small pool of shared resources such as
caches, 64- and 32-bit local registers, and some shared memory.
It is ultimately these shared resources and the number of thread
blocks that limit the number of compute threads the GPU can process in
parallel at any one point in time.

\begin{figure}[!htbp]
  \centering
  \begin{subfigure}[t]{\textwidth}
      \centering
      \includegraphics[width=\textwidth]{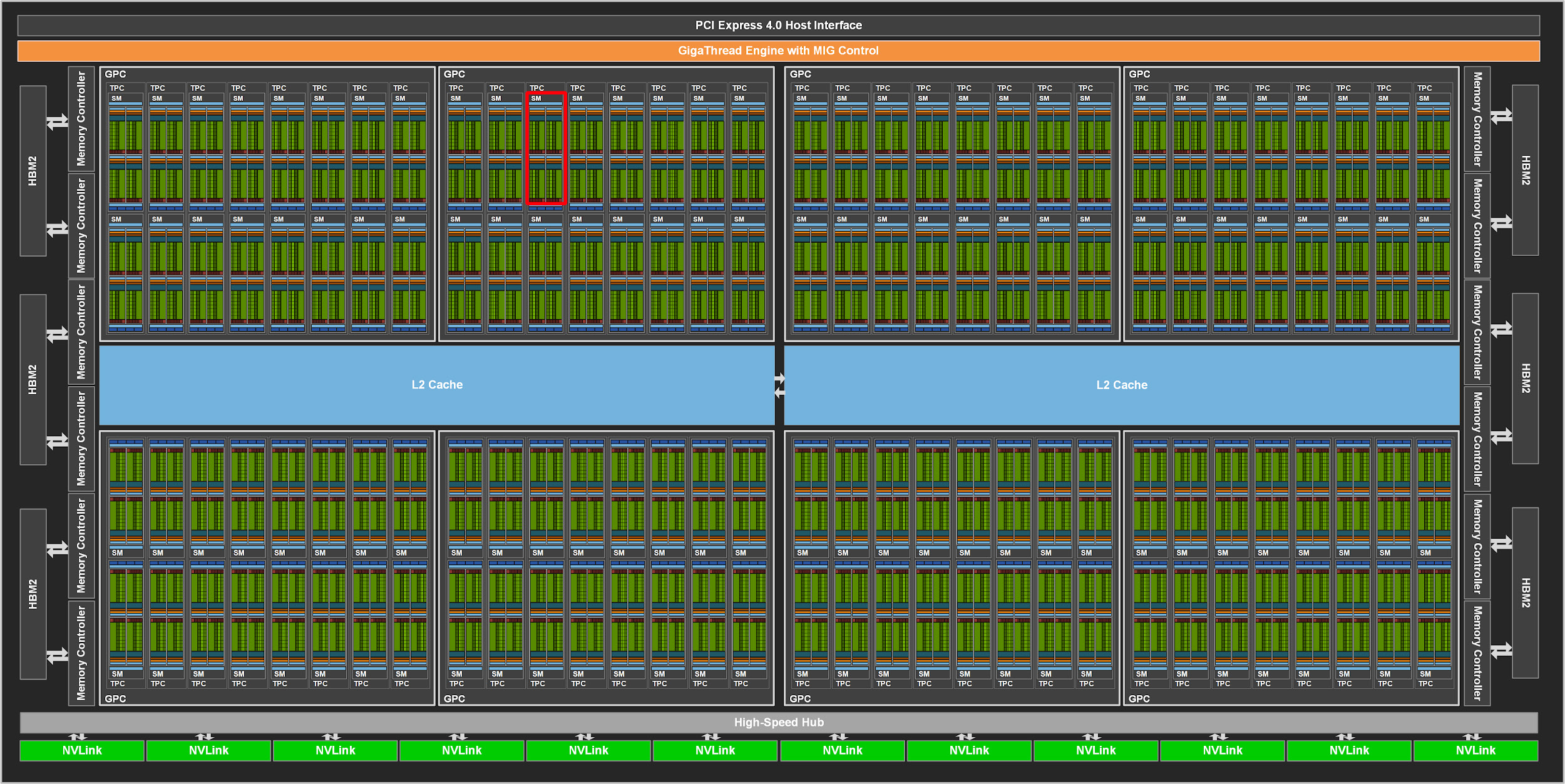}
      \caption{Diagram of a complete NVIDIA A100 GPU die \cite{NVA100_paper}.
      The red rectangle, added by us for clarity, shows the location 
      of a single SM\@.}
  \end{subfigure}

  \begin{subfigure}[t]{0.5\textwidth}
    \centering
    \includegraphics[width=\textwidth]{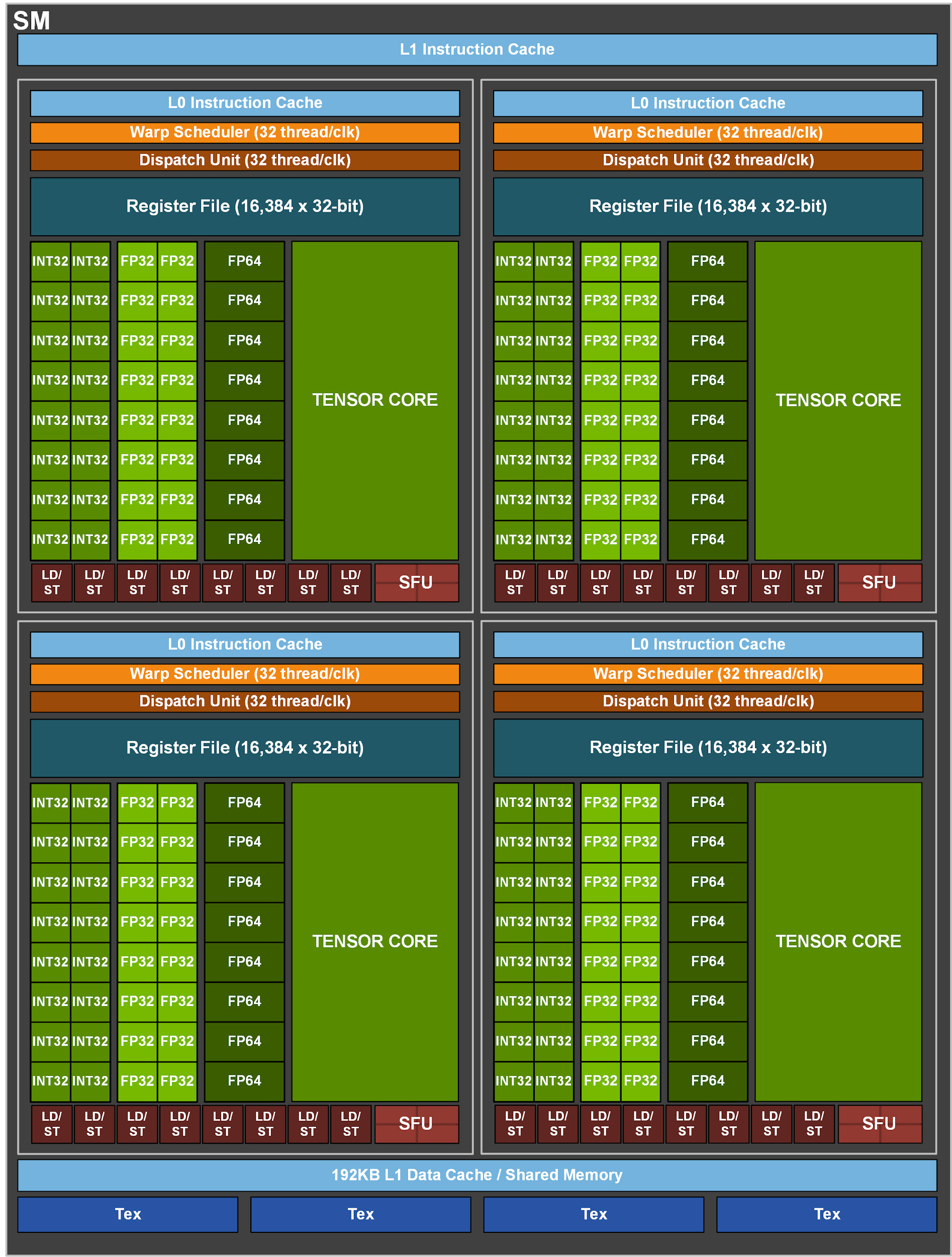}
    \caption{Exploded view of a single SM from an NVIDIA A100 
    GPU \cite{NVA100_paper}.}
\end{subfigure}
  \caption{Schematic of an NVIDIA A100 die and an individual SM\@.}\label{fig:SM}
\end{figure}

Slightly confusingly, each device manufacturer has its own naming
convention for these components, and these naming conventions are not
always directly comparable.
Figure \ref{fig:SM} shows a schematic of an NVIDIA A100 GPU die and an
individual SM, to help clarify the terminology.

AMD refers to its compute clusters as ``compute units'' and the
underlying GPU cores are referred to as ``stream processors.''
Meanwhile NVIDIA's terminology is more complex.
They refer to compute clusters as ``streaming multiprocessors.''
However, they also use various marketing terms, all of which end in
``core'' and, crucially, none of which refer to the actual GPU cores
themselves.
The most commonly misused of these is ``CUDA cores,'' which refer to
the floating-point vector processors in each cluster, not the GPU
cores.
They also use the term ``RT cores'' to refer to separate hardware used
for ray-tracing and ``Tensor cores'' to refer to hardware for matrix
operations used extensively in artificial intelligence (AI) workloads.
These cores are not utilized by CASINO due to their lack of support
for DP arithmetic.

Another terminology worth mentioning at this stage is that used by the
various programming frameworks themselves.
In terms of software, GPU APIs use various terms to express the
hierarchical parallelism of the underlying hardware.
The exact mapping, however, is not always one-to-one, as the compiler
transparently handles the mapping of the logical parallelism to the
actual hardware.
It is however helpful to discuss these terms as they are used in the
context of different GPU programming languages.

NVIDIA's programming language CUDA uses the terms ``warps'' and
``thread blocks.''
In simple terms a warp is the smallest compute unit the programmer has
access to, a set of 32 threads, all of which work on the same
program instruction in lockstep.
From a hardware perspective warps can be thought of as instructions
running on individual vector processors within a GPU core.

A single compute cluster can run a single instruction over multiple
warps, grouped into thread blocks using the scheduling hardware.
At the time of writing the current and previous two generations of
NVIDIA GPUs allow each SM to handle a maximum of 64 total warps spread
across a maximum of 32 thread blocks.

The CUDA API uses CUDA grid dimensions $x$ and $y$ for indexing threads.
The compiler handles the mapping between these logical coordinates and 
the underlying hardware, so these dimensions do not necessarily need to 
be a multiple of 32. 
The compiler simply pads the instruction with zeros to map to the next
largest warp.
Thus the exact number of warps used for each instruction is not directly 
user-controllable.

AMD's programming language ROCm helpfully uses similar terminology to
CUDA for the most part, with the only differences being that
``blocks'' are called ``workgroups'' and that warps contain 64
threads.

Finally OpenACC, used by CASINO, uses similar concepts with ``gangs''
``workers'' and ``vectors''. 
However these in general are not directly comparable to the CUDA/ROCm 
terminology.
``Workers'' and ``vectors'' perform a similar role to the CUDA/ROCm
$x$ and $y$ dimensions.
However, the exact hardware mapping is handled transparently by the
compiler.
``Gangs'' also do not have a direct hardware equivalent, but can be
thought of as a group of threads that are scheduled together.
Usually each gang is mapped to a single compute cluster but in general
this is merely a suggestion from the programmer to the compiler.
Note there is also no specific OpenACC term for warps; they are just
blocks of 32 or 64 vectors, and programmers are generally advised to
choose an appropriate number of vectors (i.e., a multiple of 32) to
tune performance to whatever hardware is being used.

For these reasons, although OpenACC is hardware-agnostic by design,
there are important hardware considerations to achieving good
performance.
In this work we are focusing particularly on using NVIDIA hardware
with the NVIDIA toolkit, specifically Nsight Compute, to analyze
performance.
Therefore in the remainder of this work we will use the NVIDIA
terminology.

To help clarify terminology we provide Table \ref{tab:table1} to map
common terms.

\begin{table}[!htbp]
  \centering
  \caption{Comparison of terminology used by different GPU programming
    languages/frameworks.}\label{tab:table1}

  \begin{tabular}{lccc}
  \toprule

  NVIDIA     & AMD                           & Indicative  \\ 
  (CUDA)     & (ROCm)                       & OpenACC mapping \\ 

  \midrule

  SM         & Compute unit         & Gang \\

  CudaDim.y  & RocmDim.y     & Worker \\

  CudaDim.x  & RocmDim.x                             & Vector\\

  Warp       & Warp                     & 32/64 vectors\\   

  \bottomrule
  \end{tabular}
  \caption{Terminologies used by the different vendor APIs and a
    typical mapping to OpenACC terms.
Note that OpenACC does not guarantee any particular mapping, and this
is indicative only.
In particular, the compiler may assign multiple gangs to a single
compute cluster.
Also, an OpenACC vector is a user-defined number of threads, and often
corresponds to a small number of warps, rather than just one.}
\end{table}

\section{Use of OpenACC for offloading in CASINO \label{sec:openacc_casino}}

\subsection{Ewald interactions \label{sec:OpenACC_Ewald}}

To accelerate the evaluation of Ewald interactions we focused on
offloading Eq.\ (\ref{eq:Ewald}) to a GPU\@.
For testing we used VMC simulations of a 3D homogeneous electron gas
(HEG), with the number of particles ranging from 100 to 1630, which
spans the range of particles one would use in practical simulations.
This system was chosen primarily because we can easily scale the
system size without having to re-optimize the wave function.
The calculations were performed on an NVIDIA V100 GPU\@.
Our initial attempt consisted of parallelizing the loop over each
particle across the GPU threads in a straightforward manner.
Using the OpenACC LOOP directive we computed $v_{\rm E}$ for each
particle on its own thread in parallel.
This in turn led to us spawning between 100 and 1630
threads on the GPU, depending on the number of particles simulated,
resulting in a noticeable, if underwhelming, overall speedup of around
10--12\%.

Modern GPUs can handle many hundreds of thousands of threads
simultaneously without issue.
Therefore, while the V100 GPU is not the latest offering from NVIDIA,
we would na\"{\i}vely expect significantly better performance than
this from the underlying hardware.
The NVIDIA V100 has 84 SMs, with a maximum of 64 warps per SM\@.
Therefore, assuming the SMs have adequate resources, it has a
theoretical maximum of 172,032 concurrent threads (that is 84 SMs
running two blocks of 32 warps of 32 threads) \cite{NVVolta_guide}.

\begin{figure}[!htbp]
  \centering
  \begin{subfigure}[t]{0.75\textwidth}
      \centering
      \includegraphics[width=\textwidth]{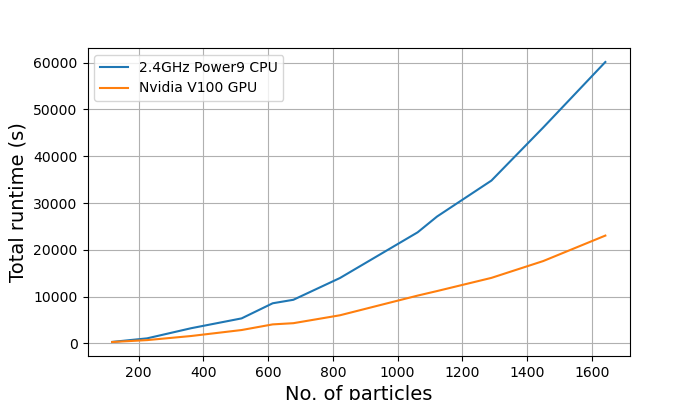}
      \caption{Total runtime vs.\ number of particles.}
  \end{subfigure}

  \begin{subfigure}[t]{0.75\textwidth}
    \centering
    \includegraphics[width=\textwidth]{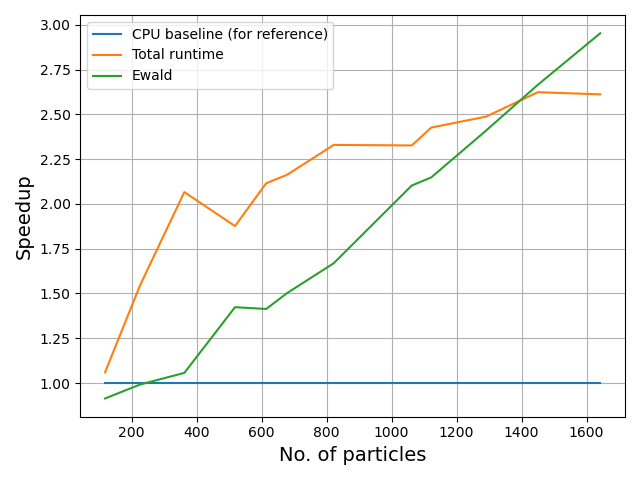}
    \caption{Speedup relative to the CPU vs.\ number of particles.}
\end{subfigure}
  \caption{Runtime comparison for VMC simulations of the 3D HEG with
    and without offloading of Ewald interactions to the GPU\@.
All runs were performed using all 32 cores of a 
POWER9 CPU @ 2.7 GHz with a single NVIDIA V100 GPU\@. \label{Ewald_figs}}
\end{figure}

Even with the maximum number of particles, it is clear we are not even
close to saturating the GPU, so our comparatively poor performance
must by caused by some other limiting factor.
Commonly this is the available shared memory or the number of
available registers.
In our case the root cause was traced, using NVIDIA's Nsight Compute
profiler, to the GPU occupancy being limited to eight thread blocks on
a single SM (512 total threads).
This limit was being enforced by the GPU hardware due to the number of
64-bit GPU registers being used, which is directly proportional to the
number of private variables used in the underlying loop.
Thus, to improve the performance, we needed to do two things:
First, reduce the number of private variables used in the loop, and
hence decrease the number of GPU registers being used.
Second, increase the number of available threads to make better use of
the underlying hardware.
This was achieved through the realization that the summations over
real and reciprocal lattices used in Eq.\ (\ref{eq:Ewald}) are
independent of one another.
Thus we can split the single, large loop over all particles into three
smaller loops: one to compute the sum over real lattice vectors, one
to compute the sum over reciprocal lattice vectors, and the final loop
to combine them at the end to get $v_{\rm E}$.
This not only reduces the number of private variables required, as the
loops are smaller, but also increases the number of available threads
as the loops over real and reciprocal space can be computed
asynchronously.
With this modification, offloading of Ewald interactions leads to an
overall speedup compared to the CPU in simulations with more than 350
particles.
Simulations with fewer particles give a less significant overall
speedup, which is likely due to the lack of work for the GPU; fewer
particles means there is less scope to use the large number of
available threads.
We also note that the fraction of run-time spent in the Ewald function
decreases linearly with particle count, at least for large enough
simulations (typically 300 particles or more).
However, since the Ewald interactions only account for 36\% of the
overall runtime in the CPU-only case, the overall effect of GPU
acceleration quickly provides diminishing returns, with the overall
speedup appearing to plateau at around 1.43 times.
The full results are shown in Fig.\ \ref{Ewald_figs}.

\subsection{Long-range two-body Jastrow terms \label{sec:u_and_p}}

For the long-range Jastrow factor discussed in Section \ref{sec:lrjas}
we focused our efforts on offloading the evaluation of the values and
derivatives of the two-body terms $u(r)$ and $p({\bf r})$ as given by
Eqs.\ (\ref{eq:u}) and (\ref{eq:p}), respectively.
As in Section \ref{sec:OpenACC_Ewald} we tested our changes in
simulations of 3D HEGs with 100 to 1930 particles performed on an
NVIDIA V100 GPU\@.
This spans the range of particles one would use for practical
simulations and was chosen primarily because we can easily scale the
system size without having to redo the wave-function optimization.
The Jastrow $u$ and $p$ terms contain random parameters, with the
cutoff length for $u$ set to the radius of the largest sphere that can
be inscribed in the Wigner-Seitz cell.

For both terms our initial attempts led to a significant slowdown of
around 3.84 times for the $u$ term and 1.28 times for the $p$ term 
with 678 particles.
A full breakdown of speedup for each of the terms can be found in
Table \ref{Jastrow_table}.
We investigated the cause of the slowdown for both terms using the
NVIDIA profiling tools Nsight Systems and Nsight Toolkit.

\begin{table}[!htbp]
  \footnotesize
  \centering
  \caption{Runtimes and speedups for offloading the $p$ and $u$
    two-body Jastrow terms in 3D HEG VMC calculations with 678
    particles.
}\label{Jastrow_table}

  \begin{tabular}{lcccc}
  \toprule

  \multirow{2}{*}{Activity} & \multirow{2}{*}{Run type} &
  \multirow{2}{*}{Runtime (s)} & Speedup & \multirow{2}{*}{\%age
    runtime} \\

& & & vs.\ CPU & \\

  \midrule

  Total runtime & CPU & 9311.46 & 1.00 & 100\%\\
  Total runtime & GPU (init.\ attempt) & 12816.79 & 0.73& 100\%\\
  Total runtime & GPU (final version) & 4307.02 & 2.16 & 100\%\\
  $P$ evaluation & CPU & 8513.60 & 1.00 & 91.43\%\\
  $P$ evaluation & GPU (init.\ attempt) & 10895.20 & 0.78 & 85.01\%\\
  $P$ evaluation & GPU (final version) & 3520.12 & 2.42 & 81.73\%\\
  $U$ evaluation & CPU & 114.95 & 1.00 & 1.23\%\\ 
  $U$ evaluation & GPU (init.\ attempt) & 442.07 & 0.26 & 5.65\%\\
  $U$ evaluation & GPU (final version) & 111.12 & 1.03 & 2.58\%\\
  \bottomrule
  \end{tabular}
\end{table}

For the $u$ term the occupancy was limited to only four warps on a
single V100 SM (for OpenACC vector length of 128, i.e., 128 concurrent
threads).
This was similar to the situation for Ewald interactions described in
Section \ref{sec:OpenACC_Ewald} in that it is a hardware-imposed limit
due to the number of 64-bit (DP) GPU registers required (which is
directly proportional to the number of private variables used in the
underlying loop).
However, unlike the Ewald interactions, our options for improving this
situation are much more limited.
We cannot break the underlying loop down any further, as each loop
iteration is its own distinct task.
The only option available was to reduce the number of 64-bit variables
used in the loop, for example by combining $x$, $y$, and $z$
components into three-element arrays.
This somewhat improved the situation.
However, the GPU performance is now merely on par with the CPU\@. 
Thus, to realize any more performance, the only option remaining is to
move to using SP arithmetic, which in effect doubles the number of
available registers.
Therefore, we have decided to keep the $u$ term on the CPU for the
production version of the code when using DP arithmetic.
Meanwhile we defer the discussion of SP performance until Section
\ref{sec:sp}.

For the $p$ term, the cause of the slowdown was somewhat more complex.
Our initial investigations revealed that the GPU hardware was limiting
the performance to a single warp on a single SM (32 threads).
Further investigations then identified this as an issue with the
amount of shared memory used by each GPU thread.
To understand the problem we must first briefly discuss the memory 
hierarchy for GPUs.
GPU threads generally have access to four levels of memory.
NVIDIA names these as L0, L1, L2, and global memory.
These can be thought of as being analogous to L1 cache, L2 cache, and
DRAM for a CPU\@.
They can be seen as light blue rectangles in Fig.\ \ref{fig:SM}.
L0 is tiny and is used only to store the current instruction being
worked on by the thread block.
The L1 cache is shared across all thread blocks within a single SM 
and is the fastest for threads to access.
However it is not very large, totalling 128 kB per SM for an NVIDIA
V100 \cite{NVVolta_guide}, and is generally only intended for
frequently accessed local and temporary variables.
The L2 cache is slower to access but is shared between all the SMs and 
is slightly larger at 6 MB \cite{NVVolta_guide}.
Finally the global memory is the slowest to access, but can be
accessed by any thread across the entire GPU and is orders of
magnitude larger at 32 GB \cite{NVVolta_guide}.
When evaluating the $p$ term, each thread needs access to three
complex arrays $e^{i{\bf b}_1\cdot{\bf r}_{ij}}$, $e^{i{\bf
    b}_2\cdot{\bf r}_{ij}}$, and $e^{i{\bf b}_3\cdot{\bf r}_{ij}}$
used in Eq.\ (\ref{eq:cosp}) to compute all the required cosines for
Eq.\ (\ref{eq:p}).
These arrays were marked as private to each thread and as such had to be 
stored in the L1 cache.
Individually the arrays are reasonably small, however, due to the
large number of threads, the arrays do not all fit into the L1 or L2
caches.
This was further compounded by OpenACC not allowing for direct mapping
of threads to the hardware, as this is handled by the compiler.
In our case the compiler was only allocating a single SM for all the 
threads in the loop, since the V100 can theoretically handle up to 2048 
threads per SM (amounting to 32 blocks of 64 threads) \cite{NVVolta_guide}.
This was ultimately the cause of the slowdown.
Each thread block was using too much shared memory.
As a result the SM was only able to schedule one thread block at a
time.
This was severely hampering the performance of the GPU as it was not
only limiting the number of parallel threads but also wasting many
CPU/GPU cycles needlessly copying data to and from shared memory.
In order to fix this, we need to keep these arrays in the global
device memory to allow the hardware to launch many more concurrent 
threads.
This was achieved by allocating three large $3 \times N$ 2D arrays.
Each thread is then given one row of each array to work on, corresponding 
to the index of the current particle.
This modification leads to a substantial performance increase of 2.42
times compared to the CPU for 678 particles (see Table
\ref{Jastrow_table}) and hence a decrease in the overall runtime of
around 53.74\%.

\begin{figure}[!htbp]
  \centering
  \begin{subfigure}[t]{0.75\textwidth}
      \centering
      \includegraphics[width=\textwidth]{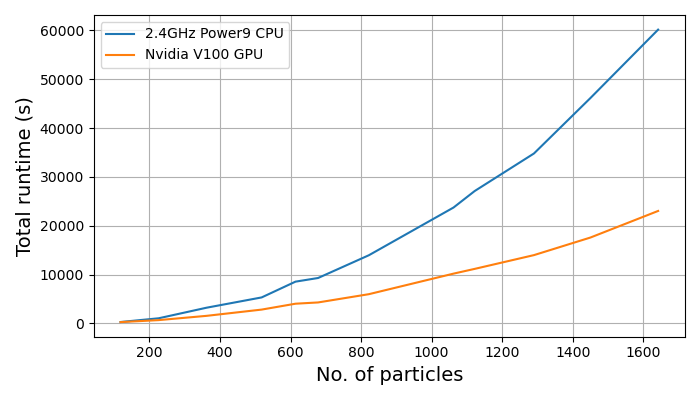}
      \caption{Runtime vs.\ number of particles.}
  \end{subfigure}
~
  \begin{subfigure}[t]{0.75\textwidth}
    \centering
    \includegraphics[width=\textwidth]{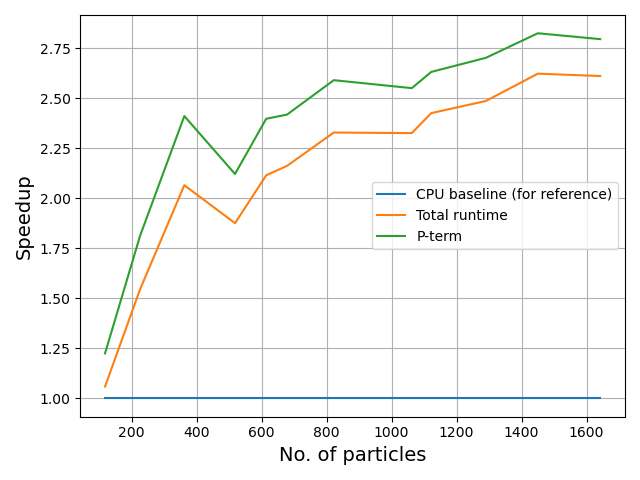}
    \caption{Speedup of different activities relative to the CPU
      vs.\ number of particles.}
\end{subfigure}
  \caption{Runtime comparison for VMC simulations of a 3D HEG with and
    without offloading of the two-body Jastrow $p$
    terms to the GPU\@.
    The results have been averaged over five runs.
    All runs were performed using all cores of a 32-core POWER9
    CPU @ 2.7 GHz with an NVIDIA V100 GPU\@.}\label{Jastrow_figs}
\end{figure}

The full results for offloading both the Ewald interactions and
Jastrow $p$ term for 100 to 1920 particles can be found in
Fig.\ \ref{Jastrow_figs}.
From this we can see that the CPU and GPU runtimes steadily diverge
from one another such that by around 1200 particles the GPU version is
roughly 2.4 times faster.
The total speedup also appears to level off as we further increase the
particle count.
This is likely a hardware-imposed bottleneck as the
memory usage increases linearly with particle count.
Therefore as particle count increases the number of threads that can
be run concurrently decreases.

One final thing to note at this stage is that the relative speedup
data (Fig.\ \ref{Jastrow_figs}b) appear quite noisy as a function of
system size.
This is due to the intrinsic variance in the runtime arising from
different random walks being taken (e.g., different numbers of moves
being accepted or rejected) in a short simulation; averaging over
multiple runs with different random seeds at each system size allows
one to reduce the apparent noise.

\subsection{Evaluation of blip orbitals \label{sec:blips}}

Since the calculations used for the evaluation of blip orbitals
consist of straightforward linear algebra, GPU offloading is likewise
straightforward.
The approach on the CPU makes use of the Level 1 and Level 2 basic
linear algebra subprograms (BLAS) libraries, specifically:
\texttt{ddot}, for performing the dot product of two DP vectors, and
\texttt{dgemv}, for calculating the DP matrix-vector operation ${\bf
  y} \mapsto \alpha A{\bf x} + \beta {\bf y}$, where $\alpha$ and
$\beta$ are scalars, ${\bf x}$ and ${\bf y}$ are vectors, and $A$ is a
matrix \cite{blas2002}.

There exist equivalent libraries for offloading BLAS calculations to a
GPU\@.
The most popular of these is cuBLAS, which is supplied by NVIDIA as
part of its CUDA software suite, and is designed to be a simple
drop-in replacement \cite{cuBLAS}.

To test the GPU performance, we simply swapped all the calls to BLAS
with their cuBLAS equivalents, while handling the data movement from
CPU to GPU using OpenACC\@.
We then performed simulations of 3D HEGs ranging from 100 to 1930
particles on an NVIDIA V100 GPU\@.
These tests, however, proved fruitless, causing a slowdown of around
five times.

\begin{figure}[!htbp]
  \centering
  \begin{subfigure}[t]{\textwidth}
      \centering
      \includegraphics[width=\textwidth]{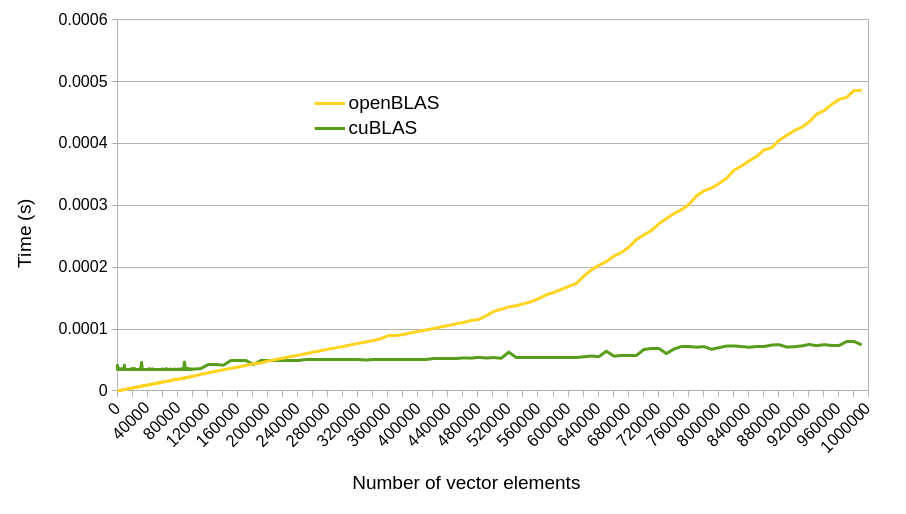}
      \caption{Runtime vs.\ number of vector elements.}
  \end{subfigure}
~
  \begin{subfigure}[t]{\textwidth}
    \centering
    \includegraphics[width=\textwidth]{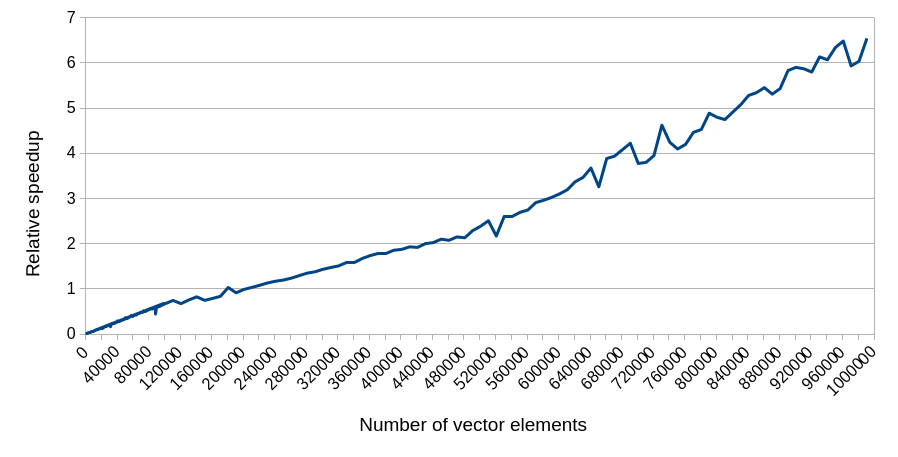}
    \caption{Speedup relative to the CPU vs.\ number of vector elements.}
\end{subfigure}
  \caption{Runtime comparison of OpenBLAS (on CPU) vs.\ cuBLAS (on
    GPU) for the \texttt{ddot} function with an increasing number of
    vector elements.
  These calculations were performed using all cores of a
  32-core POWER9 CPU @ 2.7 GHz and an NVIDIA V100
  GPU\@.}\label{Cublas_bmarks}
\end{figure}

Such results however, are not entirely surprising.
The CPU implementation of the BLAS libraries are already highly
optimized.
Thus, to overcome the overhead of the data transfer from CPU to GPU,
we ideally would need to use vectors with a very large number of
elements.
Our own benchmarks suggest the need for around 200,000 elements before
breaking even when using cuBLAS for \texttt{ddot} (see
Fig.\ \ref{Cublas_bmarks} for details).
CASINO blip-orbital evaluations make use of many small calls to BLAS,
generally using 64 element vectors and matrices.
These are not well suited for cuBLAS due to the number of data
transfers from CPU to GPU and the relatively low number of threads
utilized; hence the relatively poor GPU performance.

There is however still some scope for batching the calls to cuBLAS to
be run concurrently using CUDA streams, thus allowing us to better
utilize the GPU hardware by calculating several functions in parallel.
This approach however would require significantly more work and as
such has been left for future investigation.

\section{Use of SP arithmetic in CASINO \label{sec:sp}}

\subsection{Switching precision}

The majority of real and complex variables in CASINO are of DP kind.
The only exceptions are that (i) SP variables are used in some
low-level tasks such as timing routines, and (ii) blip coefficients
can optionally be stored as SP numbers, to reduce memory requirements.
Adopting SP throughout CASINO offers a significant speedup on a CPU
and potentially an even more significant speedup on a GPU\@.

Due to the use of Fortran, it is straightforward to switch precision.
The ``kind'' of real and complex numbers used throughout CASINO is a
parameter defined in a single module.
A small BASH script (i) changes the default kind parameter to SP and
(ii) replaces calls to DP BLAS and LAPACK routines with calls to their
SP equivalents.
The kind parameter is used to select the appropriate kind flag for
calls to MPI routines.
A few parts of the code relating to the initial setup of a
calculation, such as the evaluation of lookup tables for Ewald
interactions, are numerically sensitive and are always performed using
DP arithmetic.
Substituting the BLAS and LAPACK routines is easy because of the
rigorous naming conventions for these routines.
The apparently obvious alternative of overloading external BLAS and
LAPACK routines is nontrivial because of the use of assumed-size
arrays in these routines.
For other external libraries used by CASINO, such as Netlib's
specfun library~\cite{specfun} for special functions, we simply
overload wrapper functions in CASINO\@.

To enable the successful use of SP arithmetic it was necessary to
re-express many hardwired tolerance parameters throughout the CASINO
software in terms of the machine precision (and in one or two places
in terms of the largest and smallest representable numbers).
Fortunately Fortran provides intrinsic functions for just this
purpose.

\subsection{Accuracy of SP arithmetic in QMC calculations}

Switching from DP to SP makes surprisingly little difference to VMC
and DMC results when a blip basis set is used for the orbitals.
For example, the VMC total energy of a $2\times 2\times 2$ supercell
of bulk Si at the simulation-cell Baldereschi point using Trail-Needs
Si pseudopotentials was found to be $-7.875606(14)$ Ha per primitive
cell with SP arithmetic and $-7.875615(10)$ Ha per primitive cell with
DP arithmetic.
In these calculations, a Slater-Jastrow trial wave function was used,
with $u$, $\chi$, $f$, and $p$ terms in the Jastrow exponent and
complex Perdew-Burke-Ernzerhof orbitals represented in a blip basis
set in the Slater determinant.
The parameters in the Jastrow factor were optimized by energy
minimization with DP arithmetic.
The corresponding DMC energies at a time step of 0.04 Ha$^{-1}$ were
found to be $-7.88720(4)$ and $-7.88713(4)$ Ha per primitive cell
using SP and DP arithmetic, respectively.
Similarly excellent agreement between SP and DP QMC results is found
in a $2\times 2\times 2$ supercell of Si at the simulation-cell
$\Gamma$ point, where the trial wave function is real.
These results demonstrate that SP arithmetic is perfectly adequate for
at least some QMC calculations.

On the other hand, when an optimized Gaussian basis set is used for a
$2\times 2\times 2$ supercell of Si at the $\Gamma$ point with a
Troullier-Martins Si pseudopotential, the Slater-Jastrow VMC SP
and DP energies are $-7.88573(3)$ and $-7.88283(3)$ Ha per primitive
cell, respectively.
These differ significantly, indicating that Gaussian basis sets are
more challenging for SP arithmetic.
It was verified that this difference is not affected by reasonable
changes to the tolerance controlling the distances beyond which
Gaussian basis functions are truncated to zero.
The difference in SP and DP total energies is equal to the
corresponding difference in kinetic energies, while the potential
energies are in statistical agreement, implying that the evaluation of
the Laplacians of orbitals expanded in a Gaussian basis set is the
part that is particularly sensitive to the choice of precision.

Furthermore, even when a blip basis set is used, wave-function
optimization is less effective with SP arithmetic than with DP
arithmetic.
Wave function optimization is usually computationally cheaper than
DMC, but is the most technically challenging aspect of QMC work and
the most costly in terms of human time.
Re-optimization by energy minimization with SP arithmetic does not
significantly alter or spoil the previously optimized trial wave
function used in the results quoted above for a $2\times 2\times 2$
supercell of Si at the Baldereschi point.
However, optimization of the parameters in that trial wave function
from scratch with SP arithmetic yields a VMC energy of $-7.86165(9)$
Ha per primitive cell, which is significantly higher than the VMC
energy of $-7.875606(14)$ Ha per primitive cell obtained using energy
minimization with DP arithmetic.

In summary, although SP arithmetic can safely be used for many types
of QMC calculation, there exist important use-cases in which SP
arithmetic is problematic.
In the present work we are primarily concerned with the speedup that
can be obtained through the use of SP arithmetic; we will not
investigate the accuracy of such calculations further.

\subsection{Speedups due to offloading with SP arithmetic}

Table \ref{Sp_Table} shows the speedup resulting from offloading a 3D
HEG VMC simulation using SP arithmetic for 678 and 1290
particles respectively.
These have been broken down into contributions from the Ewald
interactions as well as the Jastrow $u$ and $p$ terms.
From this we can see that both the Ewald interactions and Jastrow $p$
term show a similar performance improvement to the DP case.

\begin{table}[!htbp]
  \centering
  \caption{Runtimes and speedups for offloading the $p$ and $u$
    two-body Jastrow terms and Ewald interactions in VMC simulations
    of 3D HEGs with $N=678$ and 1290 particles.
SP arithmetic was used.}\label{Sp_Table}
  \begin{tabular}{lcccc}
  \toprule
  Activity & Run type & $N$ & Runtime (s) & Speedup vs.\ CPU \\ 
  \midrule
  Total runtime & CPU & 678 & 876.90 & 1.00 \\
  Total runtime & GPU & 678 & 346.71 & 2.53 \\
  $P$ eval. & CPU & 678 & 545.44 & 1.00 \\
  $P$ eval. & GPU & 678 & 193.11 & 2.82 \\
  $U$ eval. & CPU & 678 & 4.31 & 1.00 \\ 
  $U$ eval. & GPU & 678 & 3.88 & 1.11 \\  
  Ewald eval. & CPU & 678 & 20.79 & 1.00 \\ 
  Ewald eval. & GPU & 678 & 11.63& 1.79 \\
  Total runtime & CPU & 1290 & 3225.20 & 1.00 \\
  Total runtime & GPU & 1290 & 958.58 & 3.36 \\
  $P$ eval. & CPU & 1290 & 1958.12 & 1.00 \\
  $P$ eval. & GPU & 1290 & 620.70 & 3.15 \\
  $U$ eval. & CPU & 1290 & 8.75 & 1.00 \\ 
  $U$ eval. & GPU & 1290 & 7.65 & 1.14 \\  
  Ewald & CPU & 1290 & 78.10 & 1.00 \\ 
  Ewald & GPU & 1290 & 21.53 & 3.63 \\
  \bottomrule
  \end{tabular}
\end{table} 

From this we can see that both the Ewald interactions and Jastrow $p$
term see a similar performance improvement as the DP case, leading to
speedups of 2.82 and 1.79 times, respectively, for 678 and 1290 particles.

Unfortunately, despite doubling the number of available registers, the
$u$ term still sees only a small improvement of 1.11 and 1.14 times 
respectively.
Further investigations revealed that the GPU kernel used in the
computation was experiencing significant ``warp stall,'' meaning that
the SM is waiting multiple cycles for the current instruction to
complete.
This is indicative that there is simply too much work being performed
per thread.
Ideally for best performance on a GPU we want many threads each
performing a small amount of work.
This however, is not the case for the $u$ term as each iteration of
the corresponding loop performs significant work which cannot easily
be broken down into smaller parallel tasks.
As such, we conclude that offloading the $u$ term to GPU is simply not
worthwhile with present hardware and this has not been enabled in the 
release version of the code.
However, given the industry trend towards AI (which heavily favors the
use of SP and even half-precision arithmetic) a significant
performance improvement for SP workloads is not out of the question,
so this may be worth revisiting in future with newer hardware.

\begin{figure}[!htbp]
  \centering
  \begin{subfigure}[t]{0.85\textwidth}
      \centering
      \includegraphics[width=\textwidth]{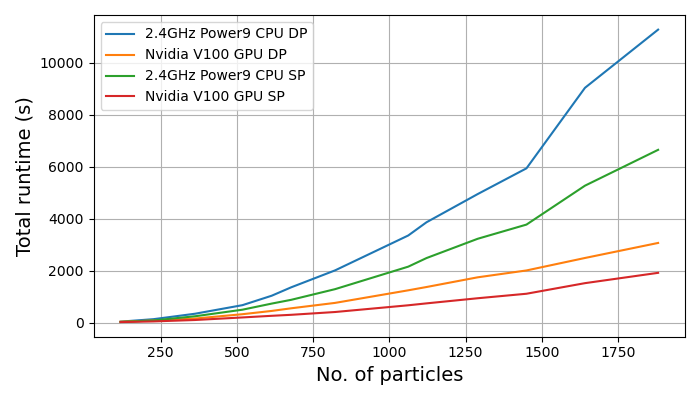}
      \caption{Runtime vs.\ number of particles.}
  \end{subfigure}

  \begin{subfigure}[b]{0.48\textwidth}
    \centering
    \includegraphics[width=\textwidth]{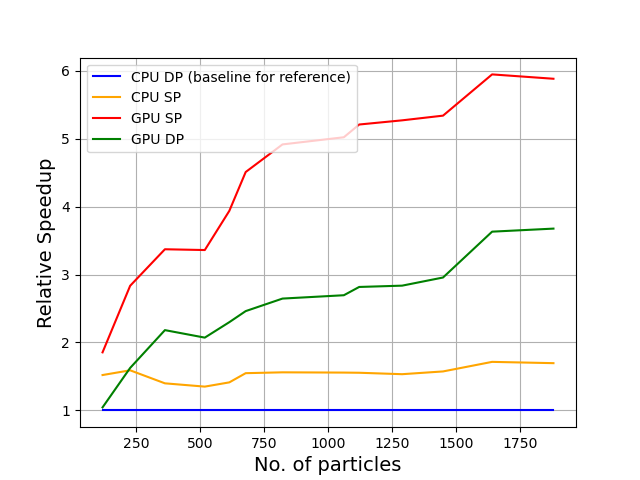}
    \caption{Speedup relative to DP CPU vs.\ number of particles.}
\end{subfigure}
~
\begin{subfigure}[b]{0.48\textwidth}
  \centering
\includegraphics[width=\textwidth]{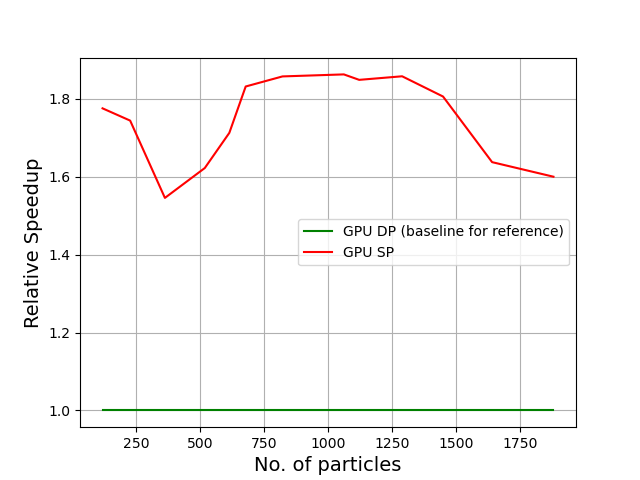}
  \caption{Speedup relative to DP GPU vs.\ number of particles.}
\end{subfigure}
  \caption{Runtime comparison for VMC simulations of a 3D HEG with DP
    and SP arithmetic averaged over five runs.
All runs were performed using all cores of a 32-core POWER9 CPU @
2.7 GHz with an NVIDIA V100 GPU\@.}\label{SP_figs1}
\end{figure}

The full results for offloading both the Ewald interactions and
Jastrow $p$ term for 100 to 1920 particles can be found in
Fig.\ \ref{SP_figs1}.
For the CPU case we can see that the use of SP arithmetic leads to a
performance improvement of around 1.5 times.
This is broadly in line with what we would expect from the POWER9
hardware used, with the smaller datatype allowing for improved CPU
cache performance, reduced memory bandwidth, and more efficient CPU
vectorization \cite{Aart2018}.
The SP GPU case also sees a significant performance improvement, being
between 1.9 and 6 times faster than the DP CPU version, and between
1.5 and 1.8 times faster than the DP GPU case.
GPUs are primarily intended for work with computer graphics, where
operations involving SP and even half-precision are commonplace;
indeed on consumer cards and some cheaper server cards it is common to
have more SP floating-point units than DP ones by a ratio of as much
as 30:1.
Therefore the large performance increases we see here are unsurprising as GPU
hardware has been designed and optimized for performance with SP
arithmetic.

\section{Performance on an NVIDIA H100}

So far we have only discussed performance on an NVIDIA V100 GPU\@.
This particular GPU, whilst still performant, is now somewhat dated,
having been released in 2017.
Therefore we have also tested the performance of CASINO on the NVIDIA
H100 GPU\@.
This GPU is based on the NVIDIA Hopper architecture and is two
generations ahead of the V100.
The H100 is expected to be significantly faster than the V100, with a
reported 2.5 times increase in peak performance for DP operations
\cite{NVVolta_guide,NVHopper_guide}.

To investigate the performance of the H100, we used the NVIDIA
Grace-Hopper GH200 \textcolor{purple}{s}uperchip allocation on Bede.
This consists of a 72 core NVIDIA Grace 64-bit ARM-based CPU @ 3.483
GHz with a 96 GB NVIDIA H100 GPU\@.

\begin{figure}[!htbp]
  \centering
  \begin{subfigure}[t]{0.85\textwidth}
      \centering
      \includegraphics[width=\textwidth]{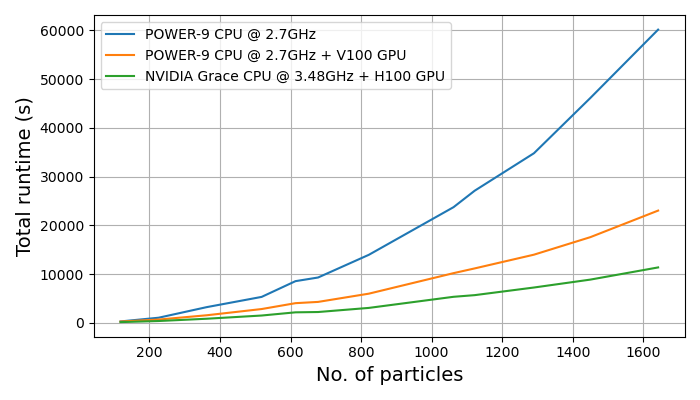}
      \caption{Runtime vs.\ number of particles.}
  \end{subfigure}

  \begin{subfigure}[b]{0.85\textwidth}
    \centering
    \includegraphics[width=\textwidth]{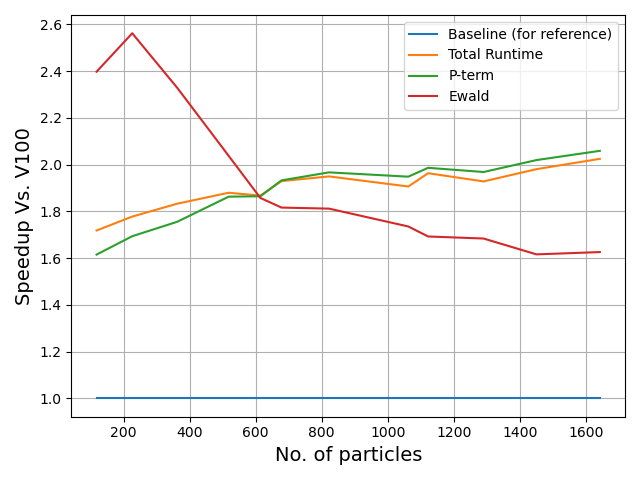}
    \caption{Speedup of GPU accelerated sections relative NVIDIA V100 
    vs.\ number of particles.}
\end{subfigure}

  \caption{Runtime comparison for VMC simulations of a 3D HEG\@.
Runs were performed using a 32-core POWER9 CPU @
2.7 GHz paired with an NVIDIA V100 GPU and an NVIDIA Grace 72-core arm64 CPU 
@ 3.483 GHz paired with an NVIDIA H100 GPU\@.}\label{H100_figs1}
\end{figure}

For our tests we have studied the 3D HEG with 118 to 1640 particles
and DP arithmetic.
The results of these tests can be seen in Fig.\ \ref{H100_figs1}.

We note at this stage that a direct comparison of performance vs.\ the
CPU is essentially meaningless as the CPU used in the Grace-Hopper
system is an ARM-based CPU, which is not designed for
compute-intensive workloads and as such is not performant for QMC
calculations.

Hence we will only compare the performance of the H100 with the V100
for the sections of code which we have accelerated using the GPU,
i.e., Ewald interactions and the Jastrow $p$ term.

From this we can see that for both the Ewald interactions and Jastrow
$p$ term the H100 is significantly faster.
For the $p$ term the H100 is around 1.6 times faster for 118
particles and the speedup relative to the V100 slowly rises with
particle count to around 2.1 times faster for 1642 particles.
However, for the Ewald interactions the H100 is 2.4 times faster than
the V100 for 118 particles, after which the speedup rises and then
peaks at 2.55 times faster for 226 particles, before slowly decreasing
to 1.61 times at 1640 particles.
All these performance increases are in line with the expected
performance increase for DP operations.

The initially higher performance for Ewald interactions at low
particle counts is likely due to the lack of available work for the
GPU\@.
As the number of particles increases, so does the number of threads,
and thus the kernel launch time.
For the V100 this effect was mostly obscured by the data transfer
time, which is in general an order of magnitude larger than the kernel
launch time.

However, in the case of the H100, the Grace-Hopper superchip grants
both the CPU and the GPU direct access to each other's memory space.
Therefore the data transfer times are essentially instantaneous, and
the effect of the increased kernel launch times is much more readily
visible, with the 2.4 to 2.55 times speedup being essentially in line
with Nvidia's own estimates for the theoretical maximum performance
uplift of 2.6 times \cite{NVA100_paper}.

\section{Further examples and discussion \label{sec:results}}

\subsection{2D HEG code performance}

CASINO is a complex codebase and, while the 3D HEG is useful for
testing and gauging the general offloading performance, it is lacking
in two main areas.
Firstly it is a 3D system, so it does not show how GPU offloading
affects 2D systems, which are a significant use-case and which use a
slightly different code path due to various simplifications and
optimizations.
Secondly and more importantly, the HEG does not include the
complications that arise from the presence of atomic cores in
\textit{ab initio} QMC calculations, such as nontrivial
single-particle orbitals, pseudopotentials, and inhomogeneous Jastrow
terms.
These more complex calculations are less dependent on homogeneous
two-body Jastrow factors and Ewald interactions.
We therefore study two final examples: a 2D HEG, to demonstrate 2D
code performance; and a $9 \times 9$ supercell of monolayer hexagonal
boron nitride (hBN), to demonstrate the performance for a more
realistic use-case of CASINO\@.

\begin{figure}[!htbp]
  \centering
  \begin{subfigure}[t]{0.75\textwidth}
      \centering
      \includegraphics[width=\textwidth]{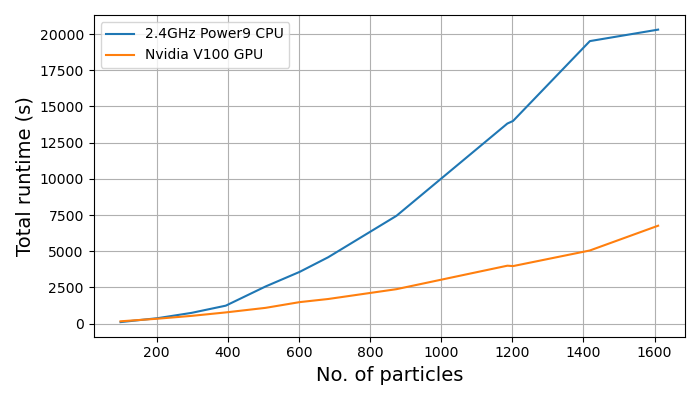}
      \caption{Runtime vs.\ number of particles.}
  \end{subfigure}
~
  \begin{subfigure}[t]{0.75\textwidth}
    \centering
    \includegraphics[width=\textwidth]{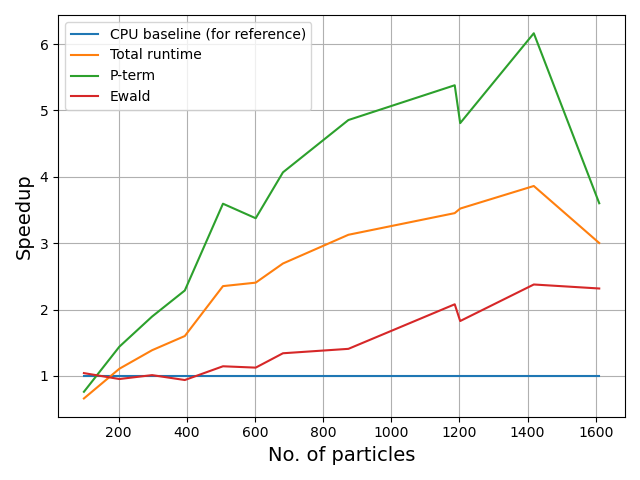}
    \caption{Speedup relative to the CPU vs.\ number of particles.}
\end{subfigure}

  \caption{Runtime comparison for VMC simulations of a 2D HEG, with
    and without offloading of the two-body Jastrow $p$ term to the GPU,
    averaged over five runs.
All runs were performed using all cores of a 32-core POWER9 CPU @
2.7 GHz with an NVIDIA V100 GPU\@.}\label{2D_figs}
\end{figure}

For our 2D test we use an equivalent example to the 3D case: a 2D
HEG with between 100 and 1642 particles.
The calculations were again performed on a POWER9 CPU with an NVIDIA
V100 GPU\@ and were averaged over five runs.
The results of these calculations can be see in Fig.\ \ref{2D_figs}.
Unlike the 3D case the 2D starts out fractionally slower on the GPU,
likely due to the low number of threads.
However, this quickly changes as the system size increases, with
the GPU version being around 3.5 times faster for 1200 particles.
The speedup for the 2D HEG appears to level off with system size, as
in the 3D case, although noticeably more slowly.
The speedup for the Ewald interactions increases at a much lower rate
than in the 3D case, and appears less linear with system size.
We also note that offloading the Ewald interaction at 98 particles
appears to show a small speedup of 1.1 times.
In reality, however, this is an artifact of the runtimes being small
and very similar to one another.
For the CPU version, the Ewald interactions took an average of 0.21
seconds with a standard deviation of 0.04 seconds, compared with 0.19
seconds with a standard deviation of 0.03 seconds for the GPU version.
This means the values were within one standard deviation of each
other, and as such this ``phantom'' speedup can be regarded as
statistically insignificant.

\subsection{Hexagonal boron nitride}

Our second example is a $9 \times 9 \times 1$ supercell of hexagonal boron 
nitride.

\begin{figure}[!htbp]
  \centering
  \begin{subfigure}[t]{0.75\textwidth}
      \centering
      \includegraphics[width=\textwidth]{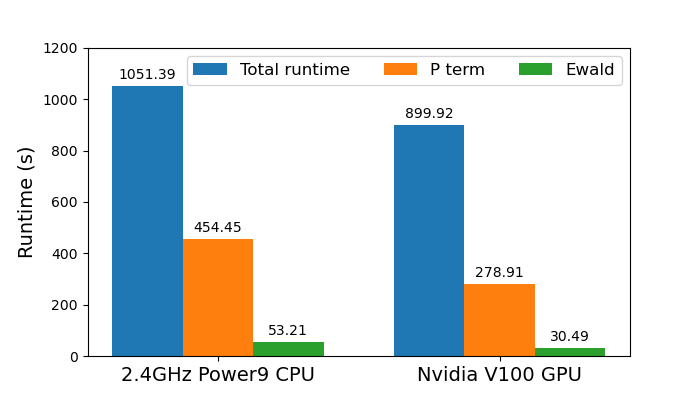}
      \caption{Total runtime and times spent computing Ewald interactions and
        the Jastrow $p$ term.}
  \end{subfigure}
  \begin{subfigure}[t]{0.75\textwidth}
    \centering
    \includegraphics[width=\textwidth]{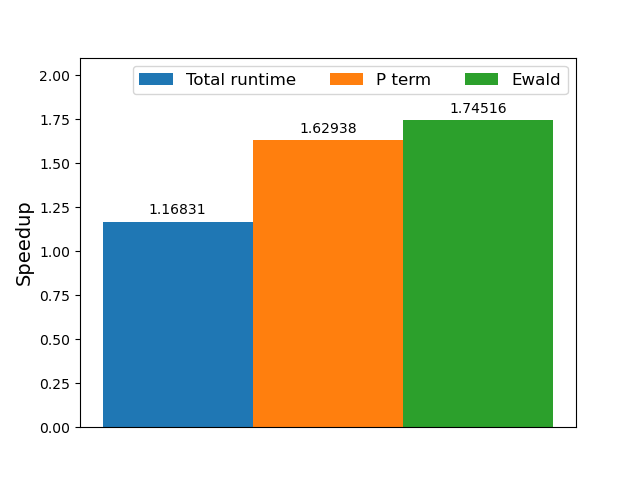}
    \caption{Speedup relative to the CPU\@.}
\end{subfigure}
  \caption{Runtime comparison for a VMC simulation of a $9 \times 9$
    supercell of monolayer hBN\@.
All runs were performed using all cores of a 32-core POWER9 CPU @
2.7 GHz with an NVIDIA V100 GPU\@.}\label{BN_figs}
\end{figure}
Figure \ref{BN_figs} shows the total runtime of the simulation,
alongside the time spent computing the Ewald interactions and the
Jastrow $p$ term.
From this we can see that, while we do see significant speedups in the
computation of both the Ewald interactions and the Jastrow $p$ term,
we do not see as significant an improvement for the overall runtime,
because both sections take up a smaller fraction of the overall
runtime than they do in the HEG\@.

\section{Conclusions\label{sec:conclusions}}

We have investigated the speedup to the CASINO code that can result
from using OpenACC to offload various aspects of a quantum Monte Carlo
calculation with a Slater-Jastrow trial wave function to a GPU\@.
For electron gases in two and three dimensions we find that, on a
POWER9 CPU with an NVIDIA V100 GPU, offloading the evaluation of Ewald
interactions is beneficial when more than about 350 electrons are
present; with 1000 electrons, offloading accelerates the evaluation of
Ewald interactions by a factor of $\sim 2.5$.
Likewise, offloading the computation of long-range two-body terms in
the Jastrow correlation factor is beneficial with more than about 150
electrons, speeding up Jastrow factor evaluation by a factor of
(again) $\sim 2.5$ with 1000 electrons.
The use of SP arithmetic increases the speedup due to
offloading to a factor of about three (on top of the speedup of about
1.3 to the CPU version due to the use of SP)\@.
In electron gases, the evaluation of two-body Jastrow terms and Ewald
interactions account for a significant fraction of the overall
runtime, and so these accelerations translate into a similar speedup
for the entire calculation.
\textcolor{purple}{A speedup of $2.5$ reduces random errors on a QMC
  calculation by a factor of $1/\sqrt{2.5} \approx 0.63$ for a given
  computational resource.
Or, for a given error bar on the total energy and a given
computational resource, CASINO could be used to study a system that is
$2.5^{1/3} \approx 1.4$ times larger.}

For \textit{ab initio} QMC calculations, the speedups to Ewald
interactions and two-body Jastrow terms are in line with the above
results for electron gases, but there are many more contributing
factors to the total runtime, including inhomogeneous Jastrow
correlations, the evaluation of single-particle orbitals in a
localized basis set, and the evaluation of pseudopotentials.
For a $9\times 9$ supercell of monolayer hexagonal BN we find an
overall speedup of 1.17 due to offloading.

These results were obtained by comparing timings for a single process
with a single thread running on a single CPU, with and without
offloading to a single GPU\@.
\textcolor{purple}{Our approach of offloading two-body Jastrow terms
  is applicable in VMC, DMC, or reptation quantum Monte Carlo
  software, while offloading of Ewald interactions is also applicable
  in path integral Monte Carlo calculations.}
An important issue to investigate is the optimal use of GPUs on
compute nodes with multiple MPI processes competing for the same set
of GPUs.
Furthermore it would be desirable to investigate the benefits of
offloading on a range of different hardware.
Finally, this work has focussed on the kinds of QMC problem of
interest in condensed-matter physics: we have studied electron gas
models and a 2D semiconductor.
In large molecular systems, electron-electron-nucleus ($f$) Jastrow
terms are relatively important, and an obvious future development
would be to investigate the offloading of these Jastrow terms.

\section*{Acknowledgements}

This work was funded by EPSRC (Grant No.\ EP/W026775/1).
The work made use of the facilities of the N8 Centre of Excellence in
Computationally Intensive Research (N8 CIR) provided and funded by the
N8 research partnership and EPSRC (Grant No.\ EP/T022167/1).
The Centre is coordinated by the Universities of Durham, Manchester,
and York.
Computer resources were also provided by Lancaster University's High
End Computing cluster.
We acknowledge useful conversations with M.\ Pacey.

\bibliographystyle{elsarticle-num} 
\bibliography{casino_openacc}

\end{document}